\documentclass[12pt,a4paper]{article}
\usepackage{amsfonts,latexsym}
\usepackage{graphicx,color}
\usepackage{dcolumn}
\usepackage{graphicx}
\usepackage{amsmath}
\usepackage{amsfonts}
\usepackage{amssymb}

\renewcommand{\thefootnote}{\fnsymbol{footnote}}

\begin{document}

\vspace{12mm}

\begin{center}
{{{\Large {\bf Gregory-Laflamme instability of black hole\\ in Einstein-scalar-Gauss-Bonnet theories }}}}\\[10mm]

{Yun Soo Myung$^a$\footnote{e-mail address: ysmyung@inje.ac.kr} and De-Cheng Zou$^{a,b}$\footnote{e-mail address: dczou@yzu.edu.cn}}\\[8mm]

{${}^a$Institute of Basic Sciences and Department  of Computer Simulation, Inje University Gimhae 50834, Korea\\[0pt] }

{${}^b$Center for Gravitation and Cosmology and College of Physical Science and Technology, Yangzhou University, Yangzhou 225009, China\\[0pt]}
\end{center}
\vspace{2mm}

\begin{abstract}
We investigate  the stability analysis of Schwarzschild black
hole in  Einstein-scalar-Gauss-Bonnet (ESGB) theory because  the instability of Schwarzschild black hole without scalar hair implies  the Gauss-Bonnet black hole with scalar hair.
The linearized scalar equation is compared to the Lichnerowicz-Ricci tensor equation in the Einstein-Weyl gravity.
It turns out that the instability of Schwarzschild black hole in ESGB theory is interpreted  as not the tachyonic instability, but the Gregory-Laflamme instability of black string.
In the small mass  regime of $1/\lambda<1.174/r_+$, the Schwarzschild solution becomes unstable and a new branch of solution with  scalar hair  bifurcates
from the Schwarzschild one. This is very similar to  finding a newly non-Schwarzschild black hole  in  Einstein-Weyl gravity.

\end{abstract}
\vspace{5mm}

\vspace{1.5cm}

\hspace{11.5cm}
\newpage
\renewcommand{\thefootnote}{\arabic{footnote}}
\setcounter{footnote}{0}


\section{Introduction}
Recently, black hole solutions with scalar hair were found from Einstein-scalar-Gauss-Bonnet (ESGB) theories~\cite{Antoniou:2017acq,Doneva:2017bvd,Silva:2017uqg}.
These black hole solutions are quite interesting because they show evasion of  the novel no-hair theorem~\cite{Bekenstein:1995un} which updated the old no-hair theorem~\cite{Bekenstein:1972ny} by the discovery of black holes with Yang-Mills~\cite{Bizon:1990sr}, Skyrme fields~\cite{Luckock:1986tr}, or
 a conformal coupling to gravity~\cite{Bekenstein:1974sf}. The novel no-hair theorem was extended to cover the standard scalar-tensor theories~\cite{Sotiriou:2011dz} and a new form covering Galileon fields  was proposed~\cite{Hui:2012qt}.
It is worth noting that these  with scalar hair  are closely related to the instability of black hole without scalar hair.
In this theory, the instability of Schwarzschild black hole is determined solely by the linearized scalar equation where the Gauss-Bonnet term acts as an effective mass. This is so because the linearized Einstein equation is nothing but that  of the Einstein gravity, where  the Regge-Wheeler
prescription works to indicate no instability~\cite{Regge:1957td,Zerilli:1970se}.

On the other hand, a fourth-order gravity (Einstein-Weyl gravity) has  provided a non-Schwarzschild black hole solution with Ricci tensor hair~\cite{Lu:2015cqa}. Here the Ricci hair means the black hole with non-zero Ricci tensor
($\bar{R}_{\mu\nu}\not=0$), compared to zero Ricci tensor ($\bar{R}_{\mu\nu}=0$) for Schwarzschild black hole.
Making use of the trace no-hair theorem which states that  $\bar{R}=0$ is zero outside the horizon as well as $\bar{R}\to 0$
at infinity and on the inner boundary at the horizon~\cite{Stelle:2017bdu}, they have found the non-Schwarzschild black hole solution which crosses the Schwarzschild solution at the bifurcation point.
In this theory, the instability of Schwarzschild black hole was determined by solving  the Lichnerowicz equation for the linearized Ricci tensor (so-called Lichnerowicz-Ricci tensor equation)~\cite{Whitt:1985ki,Mauro:2015waa,Stelle:2017bdu}.
It has  shown that the small  black hole in Einstein-Weyl gravity  is unstable against $s(l=0)$-mode of  Ricci tensor perturbation, while the large black hole is stable ~\cite{Myung:2013doa}.  Actually, this was
confirmed  by comparing the Lichnerowicz-Ricci tensor equation  with the linearized Einstein equation around a five-dimensional black string where the Gregory-Laflamme (GL) instability appeared~\cite{Gregory:1993vy}.  In  exploring  a non-Schwarzschild black hole solution,
a static eigenfunction of Lichnerowicz
operator has two crucial roles in Einstein-Weyl gravity~\cite{Lu:2017kzi}:  a role of  perturbation away
from Schwarzschild  black hole along a non-Schwarzschild black hole  and  the other role of threshold unstable mode lying at the
edge of a domain of GL instability for  a small Schwarzschild black hole. We expect that the same thing happens to the black holes found  in ESGB theories.

In this work, we will investigate a close connection for  instability of black holes found   between  ESGB theory and  Einstein-Weyl gravity.
The linearized scalar equation around Schwarzschild black hole in ESGB theory  is surely compared to the Lichnerowicz-Ricci tensor equation around Schwarzschild black hole in the Einstein-Weyl gravity.
It turns out that the instability  of black hole in ESGB theory is interpreted  as the GL instability of black string,
 even though one uses the linearized scalar equation.
In the small mass  but not extreme curvature regime, the black hole solution becomes unstable and a new branch of solution with scalar hair  bifurcates
from the Schwarzschild black hole without scalar hair. This is very similar to  finding a  non-Schwarzschild black hole with Ricci tensor hair  in  Einstein-Weyl gravity.
Importantly, this will be interpreted as a scalar theory version of the GL instability for a black hole without hair.

\section{ESGB theory} \label{sec1}

Let us  start with the ESGB theory~\cite{Doneva:2017bvd}
\begin{equation}
S_{\rm ESGB}=\frac{1}{16 \pi}\int d^4 x\sqrt{-g}\Big[ R-2\partial_\mu \phi \partial^\mu \phi-V_\phi+\lambda^2 f(\phi) {\cal R}^2_{\rm GB}\Big],\label{Action1}
\end{equation}
where $\phi$ is the scalar field  with a potential $V_\phi$ and  a coupling function $f(\phi)$, $\lambda$ is the GB coupling constant having inverse mass dimension, and ${\cal R}^2_{\rm GB}$ is the GB term defined by
\begin{equation}
{\cal R}^2_{\rm GB}=R^2-4R_{\mu\nu}R^{\mu\nu}+R_{\mu\nu\rho\sigma}R^{\mu\nu\rho\sigma}.\label{Action2}
\end{equation}
In this work, we choose $V_\phi=0$, but not choose  a specific  form for $f(\phi)$. Instead, we would like imposing  the conditions of  $f'(\phi)|_{\phi=0}=0$ and $f''(\phi)|_{\phi=0}=1$ and thus,
it admits the expansion of $f(\phi)\approx f(0)+\phi^2/2+\cdot$.
Examples include $f(\phi)=\frac{1}{2}\phi^2$~\cite{Silva:2017uqg} and $f(\phi)=\frac{1}{12}[1-e^{-6\phi^2}]$~\cite{Doneva:2017bvd}.
Other examples appeared in Ref.~\cite{Antoniou:2017hxj}.
From the action (\ref{Action1}), we derive  the Einstein  equation
\begin{eqnarray}
 G_{\mu\nu}=2\partial _\mu \phi\partial _\nu \phi -(\partial \phi)^2g_{\mu\nu}+\Gamma_{\mu\nu}, \label{equa1}
\end{eqnarray}
where $G_{\mu\nu}=R_{\mu\nu}-(R/2)g_{\mu\nu}$ is  the Einstein tensor and  $\Gamma_{\mu\nu}$   is given by
\begin{eqnarray}
\Gamma_{\mu\nu}&=&2R\nabla_{(\mu} \Psi_{\nu)}+4\nabla^\alpha \Psi_\alpha G_{\mu\nu}- 8R_{(\mu|\alpha|}\nabla^\alpha \Psi_{\nu)} \nonumber \\
&+&4 R^{\alpha\beta}\nabla_\alpha\Psi_\beta g_{\mu\nu}
-4R^{\beta}_{~\mu\alpha\nu}\nabla^\alpha\Psi_\beta  \label{equa2}
\end{eqnarray}
with
\begin{equation}
\Psi_{\mu}=\lambda^2 \frac{df(\phi)}{d\phi} \partial_\mu \phi=\lambda^2f'(\phi)\partial_\mu \phi.
\end{equation}
On the other hand, the scalar field equation takes the form
\begin{equation}
\square \phi +\frac{\lambda^2}{4}f'(\phi) {\cal R}^2_{\rm GB}=0 \label{s-equa}.
\end{equation}

Choosing  $\phi=0$ and $f'(\phi)|_{\phi=0}=0$, one finds the Schwarzschild  solution from (\ref{equa1}) and (\ref{s-equa})
\begin{equation} \label{ansatz}
ds^2= \bar{g}_{\mu\nu}dx^\mu dx^\nu=-\Big(1-\frac{r_+}{r}\Big)dt^2+\frac{dr^2}{\Big(1-\frac{r_+}{r}\Big)}+r^2d\Omega^2_2
\end{equation}
with $r_+=2M$.
Notice that (\ref{ansatz})  indicates  the black hole solution without scalar hair.

For the stability analysis, we need the two linearized equations which describe the metric perturbation   $h_{\mu\nu}$ and scalar perturbation $\delta \phi$ propagating around (\ref{ansatz}).  They are derived by linearizing  (\ref{equa1}) and (\ref{s-equa}) as
\begin{eqnarray}
  \delta R_{\mu\nu}(h) &=& 0, \label {l-eq1}\\
  \left(\bar{\square}+ \frac{\lambda^2}{4}\bar{{\cal R}}^2_{\rm GB}\right)\delta \phi&=& 0. \label{l-eq2}
\end{eqnarray}
Here the overbar( $\bar{}$ ) denotes computation based on the background spacetime (\ref{ansatz}).
Importantly,  we note that ``$-\frac{\lambda^2}{4}\bar{{\cal R}}^2_{\rm GB}$" plays a role of
not a mass $\tilde{m}^2$ but an effective mass $\tilde{m}^2_{\rm eff}$  for $\delta \phi$  because it depends on $r$.
In this sense, the GB coupling term is quite different from a negative mass term like $V_\phi=-m_{\rm T}^2\phi^2/2$ with $m^2_{\rm T}>0$ in the tachyonic scalar theory.

For comparison, we would like to mention the Lichnerowicz-Ricci tensor equation around the Schwarzschild black hole in
the Einstein-Weyl gravity~\cite{Whitt:1985ki,Myung:2013doa,Mauro:2015waa,Stelle:2017bdu}
\begin{equation}\label{Action-EW}
S_{\rm EW}=\int d^4x \sqrt{-g} \Big[\gamma R-\alpha C_{\mu\nu\rho\sigma}C^{\mu\nu\rho\sigma}\Big]
\end{equation}
where $C_{\mu\nu\rho\sigma}$ is the Weyl tensor. This theory implies the trace no-hair theorem of $R=0$.
Considering the Schwarzschild black hole (\ref{ansatz}) and linearizing the Einstein equation leads to
the Lichnerowicz-Ricci tensor equation for the traceless and transverse Ricci tensor $\delta R_{\mu\nu}$  as
\begin{equation}\label{EOM9}
\Big(\Delta_{\rm L}+m^2 \Big) \delta R_{\mu\nu}=0,~~~m^2=\frac{\gamma}{2\alpha},
\end{equation}
where the Lichnerowicz operator on the Schwarzschild background is given by
\begin{equation} \label{lichnero}
\Delta_{\rm L} \delta R_{\mu\nu}=-\bar{\square}\delta R_{\mu\nu}-2\bar{R}_{\mu\rho\nu\sigma}\delta R^{\rho\sigma}.
\end{equation}
Here, we emphasize that Eq.(\ref{EOM9}) is the tensor counterpart  to the linearized scalar equation (\ref{l-eq2}).
Actually, Eq.(\ref{EOM9}) describes a massive spin-2 mode ($\delta R_{\mu\nu}$) with mass $m$  propagating on the black hole background.
Rewriting Eq.(\ref{EOM9}) as $(2\alpha \Delta_{\rm L}+ \gamma) \delta R_{\mu\nu}=0$, one may recover the linearized Einstein equation $\delta R_{\mu\nu}=0$ in the limit of $\alpha\to0$.

\section{Instability of  black hole without scalar hair}

Performing the stability analysis of the black hole,
one uses firstly the linearized  Einstein equation (\ref{l-eq1}).
It turned  out  that the black hole is  stable when  making use of  the Regge-Wheeler
prescription~\cite{Regge:1957td,Zerilli:1970se}. In
this case,  a massless spin-2 mode starts with  $l=2$.

Now, let us consider the linearized scalar equation (\ref{s-equa}).
Considering
\begin{equation} \label{scalar-sp}
\delta \phi(t,r,\theta,\varphi)=\frac{u(r)}{r}e^{-i\omega t}Y_{lm}(\theta,\varphi),
\end{equation}
and introducing a tortoise coordinate $r_*=r+r_{+}\ln(r/r_+-1)$ defined by $dr_*=dr/(1-r_+/r) $, the radial equation of (\ref{l-eq2}) leads to the Schr\"{o}dinger-type equation
\begin{equation}
\frac{d^2u}{dr_*^2}+\Big[\omega^2-V(r)\Big]u(r)=0,
\end{equation}
where the potential $V(r)$ is given by
\begin{equation} \label{pot-c}
V(r)=\Big(1-\frac{2M}{r}\Big)\Big[\frac{2M}{r^3}+\frac{l(l+1)}{r^2}-\frac{12\lambda^2M^2}{r^6}\Big].
\end{equation}
Moreover, introducing the negative scalar potential $V_\phi=-m^2_{\rm T} \phi^2/2$  instead of $-\lambda^2 f(\phi) {\cal R}^2_{\rm GB}$ in Eq.(\ref{Action1}),
the tachyonic scalar potential takes the form
\begin{equation} \label{tpot-c}
V_{\rm t}(r)=\Big(1-\frac{2M}{r}\Big)\Big[\frac{2M}{r^3}+\frac{l(l+1)}{r^2}-m_{\rm T}^2\Big],
\end{equation}
which induces the tachyonic instability~\cite{Silva:2017uqg} because the sufficient condition for instability ($\int^\infty_{2M} dr V_{\rm t}(r)/(1-r_+/r)<0$) is always satisfied with any mass $m^2_{\rm T}>0$. In Minkowski spacetimes, the  tachyonic scalar equation takes the form of $\ddot{\varphi}_{\bf k}(t)+ ({\bf k}^2-m^2_{\rm T}){\varphi}_{\bf k}(t)=0$, leading to an  exponentially growing solution ${\varphi}_{\bf k}(t)\sim e^{\sqrt{m^2_{\rm T}-{\bf k}^2} t}$ for $ m^2_{\rm T}>{\bf k}^2$~\cite{Felder:2001kt}. This is an  origin of instability  arisen from the tachyonic mass.

In the case of $s(l=0)$-mode, from the condition of $\int^\infty_{2M} dr V(r)/(1-r_+/r)<0$,
one may introduce  a sufficient condition of an unstable bound for a mass parameter of scalar ($1/\lambda$)~\cite{Doneva:2017bvd}
\begin{equation}
\frac{M^2}{\lambda^2}<\frac{3}{10} \Rightarrow 0<\frac{r_+}{\lambda}<1.095. \label{mass-b}
\end{equation}
However, (\ref{mass-b}) is not a necessary and sufficient condition for instability.
Observing  Fig.~1 together with $r_+=1$, one finds that   three potentials $V(r)$ have  negative regions near the horizon but they become positive after crossing the $r$-axis.
Surely, these are not types of Regge-Wheeler potentials which are positive definite outside the horizon~\cite{Regge:1957td,Zerilli:1970se}.
It might  show a new feature of instability, but the threshold of instability depends on the numerical computations.
Importantly, we find  a similar behavior from the Zerlli-type potential
\begin{eqnarray}
  V_{\rm z}(r)=\frac{(1-\frac{1}{r})}{(1+m^2r^3)^2}\Big[\frac{1}{r^3}-3m^2(4r-3)
   +3m^4r^3(2r-3)+m^6 r^6\Big]  \label{Zpot}
\end{eqnarray}
 for $s(\l=0)$ of Ricci tensor perturbation around the Schwarzschild black hole with $r_+=1$ in the Einstein-Weyl gravity (see FIG. 2 in~\cite{Lu:2017kzi}) derived from the linearized equation (\ref{EOM9}).
In addition, we would like to mention that  such potentials exist around neutral black holes (black holes without charge) in higher dimensions  and the S-deformation used to prove the stability of neutral black holes~\cite{Kodama:2003kk}. This implies that the stability analysis based on the shape of the potential is regarded as a delicate issue.
On the other hand, the tachyonic potential $V_{\rm t}(r)$ indicates a quite different behavior: it develops  positive region near the  horizon, while it
approaches $-0.04$ as $r\to \infty$ for $m_{\rm T}=0.2$. This shows clearly that the instability of black hole in ESGB theory is not just the tachyonic instability~\cite{Silva:2017uqg} because the sufficient condition for instability ($\int^\infty_{2M} dr V_{\rm t}(r)/(1-r_+/r)<0$) is always satisfied with any mass $m^2_{\rm T}>0$.

\begin{figure*}[t!]
   \centering
  \includegraphics{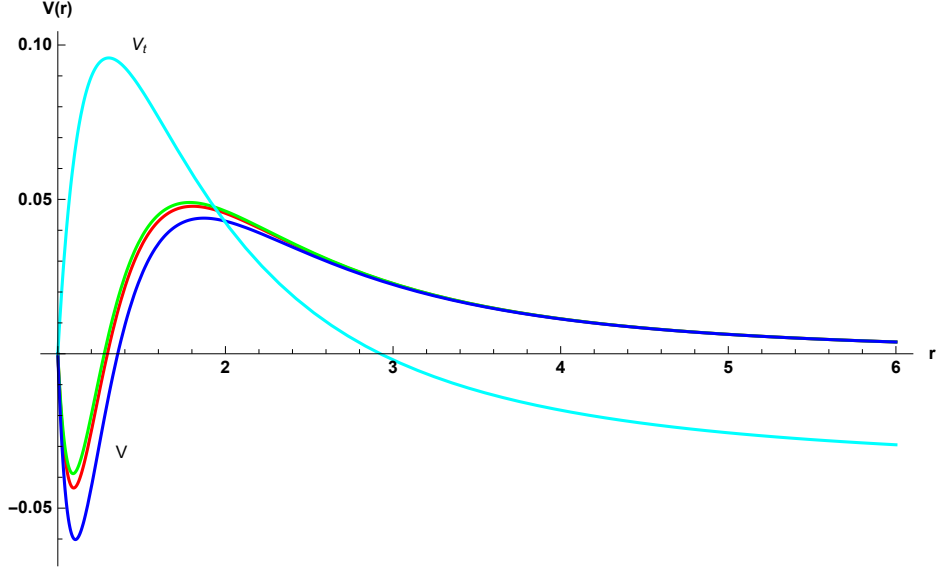}
\caption{The potentials as function of $r\in [1,\infty)$  for horizon radius $r_+=1$ and $l=0$. The blue (bottom), red (middle), and
green (top) curve represent potential  $V(r)$  of scalar for mass parameter $1/\lambda=1.095$ (sufficient condition for instability), 1.174 (threshold of instability), and 1.2 (stable case), respectively. These all have negative regions near the horizon. However, the tachyonic potential (cyan curve) $V_{\rm t}(r)$  develops  positive region near horizon while it
approaches $-0.04$ as $r\to \infty$ for $m_{\rm T}=0.2$. }
\end{figure*}

In order to determine the threshold of instability, one has to solve the second-order differential equation numerically
\begin{equation}\label{pertur-eq}
\frac{d^2u}{dr_*^2}-\Big[\Omega^2+V(r)\Big]u(r)=0,
\end{equation}
which may  allow an exponentially growing mode of  $e^{\Omega t}(\omega=i\Omega) $ as  an unstable mode.
Here we choose two boundary conditions: a normalizable
solution of $u(\infty)\sim e^{-\Omega r_*}$  at infinity  and
a solution of $u(r_+)\sim \left(r-r_+\right)^{\Omega r_+}$  near the horizon.
By observing  Fig.~2 together with $r_+=1,2,3$, we read off the
unstable bound for scalar mass parameter ($1/\lambda$) as
\begin{equation}
 0<\frac{1}{\lambda}<\Big(\frac{1}{\lambda}\Big)^{\rm th}\approx\frac{1.174}{r_+} \label{mass-c}
\end{equation}
which implies that   the threshold of instability is located at $r_+=r_c\approx 1.174$ which is greater than 1.095 (sufficient condition for instability).
This corresponds to the  bifurcation point.
Choosing $\lambda=1$, the Schwarzschild black hole will be unstable if its horizon radius satisfies the bound
\begin{equation}
r_+<r_c\approx1.174. \label{ubhb1}
\end{equation}
This implies that the Schwarzschild black hole whose radius is less than the critical radius at the bifurcation point becomes unstable, whereas
the black hole whose radius exceeds  the critical radius at the bifurcation point is stable.

\begin{figure*}[t!]
   \centering
  \includegraphics{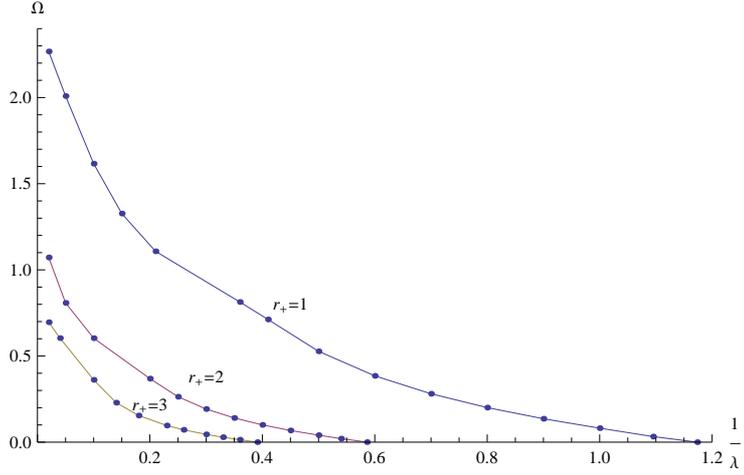}
\caption{$\Omega$ graphs as function of mass parameter $1/\lambda$ for  small
black holes of  $r_+=1,2,3$. Here, one reads off the thresholds of instability $(1/\lambda)^{\rm th}$ from the
points that curves of $\Omega$ intersect the positive $\frac{1}{\lambda}$-axis: $(1/\lambda)^{\rm th}\approx$1.174, 0.587, 0.294.
The instability range decreases as the horizon radius increases. }
\end{figure*}

\begin{figure*}[t!]
   \centering
   \includegraphics{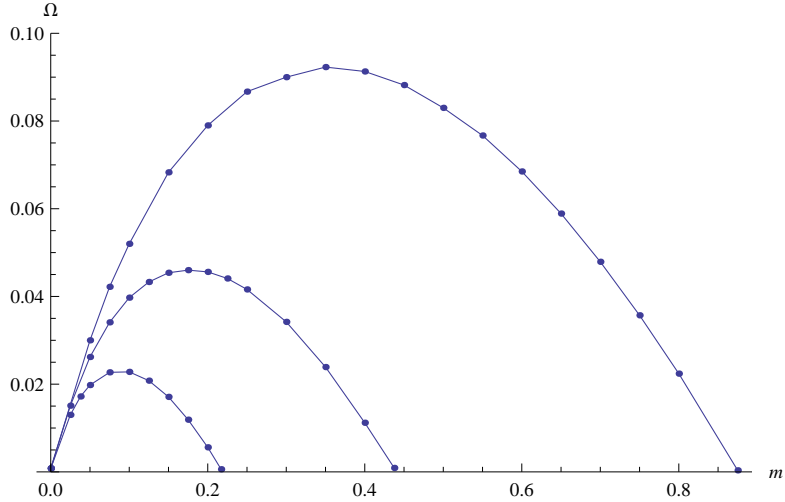}
\caption{Plots of unstable modes ($\bullet$) on three curves with the horizon radii $r_+=1,2,4$.
The $y(x)$-axis denote $\Omega$ in $e^{\Omega t}$ (mass $m$ of massive spin-2 mode). The smallest curve represents
$r_+=4$, the medium denotes $r_+=2$, and the largest one shows
$r_+=1$. Here the thresholds of instability are located at $m^{\rm th}\approx$0.876,~0438,~0.219 which means that the instability region is smaller and smaller as the horizon radius is large and larger.}
\end{figure*}

On the other hand, from Fig. 3 based on   Eq.(\ref{EOM9}),
the  GL instability mass bound for the $s(l=0)$-mode of linearized Ricci tensor $\delta R_{tr}$ in Einstein-Weyl gravity was given by~\cite{Myung:2013doa,Stelle:2017bdu}
\begin{equation}
 0<m< m^{\rm th}\approx\frac{0.876}{r_+}. \label{EWmass-c}
\end{equation}
Here, selecting $m=1$, one finds the bound for  unstable (small) black holes~\cite{Lu:2017kzi}
\begin{equation}
r_+<r_c\approx0.876.\label{ubhb2}
\end{equation}
At this stage, we note that Fig.~2 drawn for $s$-mode of scalar perturbation is similar to FIG.1 in~\cite{Moon:2013lea} for $s$-mode of Ricci tensor perturbation around the non-rotating BTZ black hole
in the new massive gravity  where the GL instability is reinforced.
In this sense, the instability arising from the bound (\ref{mass-c}) is not  the tachyonic instability~\cite{Silva:2017uqg},
but the GL instability. We stress that the tachyonic instability means negative mass squared [$-m_{\rm T}^2$ in Eq.(\ref{tpot-c})] and thus, there is no unstable bound for (positive) mass parameter ($1/\lambda$) like as (\ref{mass-c}).
\section{Static scalar perturbation}

In this section, we wish to develop the other criterion on checking the instability bound (\ref{mass-c}).
This can be achieved by exploring  the static scalar solutions to the linearized  equation (\ref{l-eq2}) on the background of Schwarzschild black hole.
Considering the expression (\ref{scalar-sp}) with $\omega=0(\Omega=0)$,
the radial equation of (\ref{l-eq2}) for  $u(r)$ can be rewritten as
\begin{equation}\label{phi-r}
\frac{r^5(r_+-r)}{3r_+^2}u''(r)+\frac{r^4}{3r_+}u'(r)-\Big[\frac{r^3}{3r_+}+\frac{l(l+1)r^4}{3r_+^2}\Big]u(r)=\lambda^2u(r).
\end{equation}
Introducing  a new  coordinate $z=\frac{r}{r_+}[z\in [1,\infty)]$ and a new  parameter
 $\lambda_s=\frac{\lambda}{r_+}$,  Eq.(\ref{phi-r}) can be rewritten as
\begin{equation}\label{phi-z}
\frac{z^5(z-1)}{3}u''(z)+\frac{z^4}{3}u'(z)-\frac{z^4}{3}\Big[\frac{1}{z}+l(l+1)\Big]u(z)=\lambda_s^2u(z),
\end{equation}
which is independent of horizon radius $r_+$. Here we focus on obtaining the $s(l=0)$-mode solution.
Because of the absence of analytic solution, one has to find a numerical solution. For this purpose, we firstly consider the near-horizon Taylor expansion for $u(z)$ as
\begin{equation}\label{expan-phi0}
u(z)=u_{+}+u'_{+}(z-1)+\frac{u''_{+}}{2}(z-1)^2+\cdots,
\end{equation}
which  can be used to set data just outside the horizon  for a numerical
integration from the horizon to the infinity. Here we note that the coefficients $u'_+$ and $u''_+$ can be determined  in terms of
a free parameter $u_+$ as
$u'_+=(1-3\lambda_s^2)u_+$ and $u''_+=\frac{3(3\lambda_s^2+2)}{2\lambda_s^6}u_+^2$.

On the other hand, an asymptotic form of
$u(z)$ near $z=\infty$ is given by
\begin{equation}\label{expan-phi1}
u(z)=u_{\infty}+\frac{u^{(1)}}{z}+\frac{u^{(2)}}{z^2}+\cdots.
\end{equation}
Similarly, we find the relations of 
$u^{(1)}=u_{\infty}/2$ and $ u^{(2)}=u_{\infty}/3$.
\begin{figure*}[t!]
   \centering
  \includegraphics{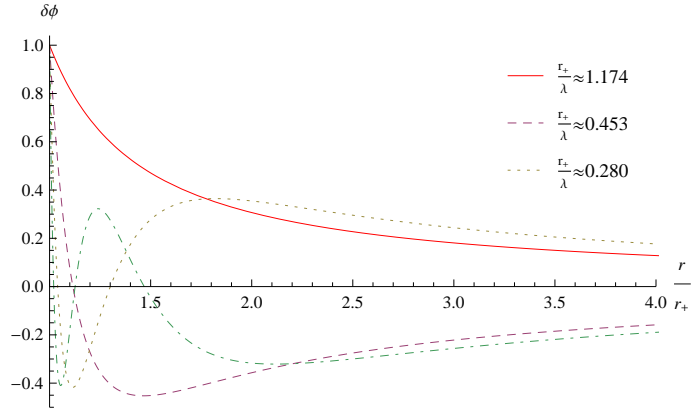}
\caption{ Graphs for scalar perturbation $\delta\phi$  as function of $r/r_+$ for $r_+/\lambda\approx1.174,~0.453,~0.280$ which correspond to the number of nodes  $n=0,~1,~2$ in scalar profiles.}
\end{figure*}

Now let us perform two integrations:  from $z=1$ to a matching point $z=z_m>1$
by imposing the ingoing wave boundary condition at the horizon  and  from $z=\infty$ to $z=z_m$ by imposing no outgoing wave at infinity.
Then, a numerical solution could be constructed by
 connecting  the near-horizon form (\ref{expan-phi0}) to the asymptotic form (\ref{expan-phi1}) together with  choosing a proper  parameter $\lambda_s$.
We obtain a discrete spectrum of parameter: $1/\lambda_s=r_+/\lambda \in [1.174,0.453,0.280,0.202,\cdots]$.
In Fig.~4, these solutions are  classified  by  order number $n=0, 1, 2, 3,\cdots$
which is  identified with the number of nodes for $\delta\phi(z)=u(z)/z$.
Here, a regular solution  to equation (\ref{pertur-eq}) with $\Omega=0$ is found only when the parameter $\lambda$
takes a specific value $r_+/\lambda\approx1.174$ (threshold of instability=the edge of domain of instability).
In other words, the $n=0$ scalar mode without zero represents a stable black hole,
while the $n=1,~2$ scalar modes  with zero  denote unstable black holes. This shows that
one may classify the black hole into   unstable and stable black holes by solving the static linearized scalar equation (\ref{phi-z}) without considering exponentially growing mode of $e^{\Omega t}$.

Finally, we note that the Schwarzschild solution without scalar hair ($\bar{\phi}=0$) is allowed for any value of $\lambda$, while the black hole solution with scalar hair ($\bar{\phi}\not=0$)
exists when $\lambda/r_+ (\lambda^2/M^2)$ belongs to a set of  scalarization bands~\cite{Silva:2017uqg}.
The key of instability  is the appearance of zeros in the scalar profiles. Actually, the discrete set corresponds to the right-end values of scalarization bands for black hole with scalar hair. This shows a close connection between instability of black hole without scalar hair and appearance of black hole with scalar hair.

\section{Discussions}
First of all, we summarize similar properties for black holes in  ESGB theory and Einstein-Weyl gravity in Table 1.
This shows a strongly  similar connection for Schwarzschild black hole between ESGB theory and Einstein-Weyl gravity even though two theories are different.

The $s$-mode of scalar perturbation around the Schwarzschild black hole induces the GL instability for the mass bound of $1/\lambda<1.174/r_+$ in ESGB theory,
while the $s$-mode of Ricci tensor perturbation around the Schwarzschild black hole induces the instability for the mass bound of $m<0.876/r_+$ in Einstein-Weyl gravity.
Also,  the instability of Schwarzschild black hole without scalar hair implies  the Gauss-Bonnet black hole with scalar hair in ESGB theory as the instability of Schwarzschild black hole without Ricci-tensor hair leads to  the non-Schwarzschild  black hole with Ricci hair in Einstein-Weyl gravity.

From this observation, we conclude that the instability of the Schwarzschild black hole in ESGB theory  is interpreted as a scalar theory version of the GL instability for a small black hole in the tensor theory of Einstein-Weyl gravity. This instability dose not belong to the tachyonic instability because the scalar potential $V(r)$ is similar to $V_{\rm z}(r)$ in (\ref{Zpot}) in Ref.\cite{Lu:2017kzi}, but it is  quite different from $V_{\rm t}(r)$ of the tachyon as was shown in Fig. 1.

 \vspace{0.5cm}
\begin{table}[h]
\resizebox{\textwidth}{!}{%
\begin{tabular}{|c|c|c|}
  \hline
  Theory & ESGB theory & Einstein-Weyl gravity\\ \hline
  Action & $S_{\rm ESGB}$ in (\ref{Action1}) & $S_{\rm EW}$ in (\ref{Action-EW})  \\ \hline
  BH without hair & SBH with $\bar{\phi}(r)=\bar{R}_{\mu\nu}=0$ &  SBH with  $\bar{R}_{\mu\nu}=0$ \\ \hline
  Linearized equation &  scalar equation (\ref{l-eq2}) &  tensor equation (\ref{EOM9}) \\ \hline
  GL instability mode & s($l=0$)-mode of $\delta\phi$ & s($l=0$)-mode of $\delta R_{\mu\nu}$ \\ \hline
  Unstable mass bound & $0<\frac{1}{\lambda}<\frac{1.174}{r_+}$ & $0<m<\frac{0.876}{r_+}$ \\ \hline
  Bifurcation point & 1.174 & 0.876 \\ \hline
  Potential & $V(r)$ in (\ref{pot-c}) &$V_{\rm z}(r)$ (\ref{Zpot}) in Ref.\cite{Lu:2017kzi} \\ \hline
  Small unstable SBH & $r_+<r_c\approx1.174$ for $\frac{1}{\lambda}=1$ &$r_+<r_c\approx0.876$ for $m=1$ \\ \hline
  BH with hair & scalar hair in Refs.\cite{Doneva:2017bvd,Silva:2017uqg} &  Ricci-tensor hair in Ref.\cite{Lu:2015cqa} \\ \hline
\end{tabular}}
\caption{Similar Properties for Schwarzschild black hole (SBH) in ESGB theory and Einstein-Weyl gravity. }
\end{table}

 \vspace{1cm}

{\bf Acknowledgments}
 \vspace{1cm}

This work was supported by the National Research Foundation of Korea (NRF) grant funded by the Korea government (MOE)
 (No. NRF-2017R1A2B4002057).

\end{document}